\documentclass[twocolumn,showpacs,preprintnumbers,amsmath,amssymb,prl,superscriptaddress,longbibliography]{revtex4-1}
\usepackage{bbm}
\usepackage{mathrsfs}
\usepackage{graphicx}
\usepackage{dcolumn}
\usepackage{bm}
\usepackage{amsmath}
\usepackage{amsfonts}
\usepackage{color}
\usepackage[colorlinks=true,citecolor=blue,anchorcolor=cyan]{hyperref}
\usepackage{floatrow}

\begin{document}

\title{Magnetic moir\'e surface states and flat chern band in topological insulators}
\author{Zhaochen Liu}
\affiliation{State Key Laboratory of Surface Physics and Department of Physics, Fudan University, Shanghai 200433, China}
\author{Huan Wang}
\affiliation{State Key Laboratory of Surface Physics and Department of Physics, Fudan University, Shanghai 200433, China}
\author{Jing Wang}
\thanks{wjingphys@fudan.edu.cn}
\affiliation{State Key Laboratory of Surface Physics and Department of Physics, Fudan University, Shanghai 200433, China}
\affiliation{Institute for Nanoelectronic Devices and Quantum Computing, Fudan University, Shanghai 200433, China}
\affiliation{Zhangjiang Fudan International Innovation Center, Fudan University, Shanghai 201210, China}

\begin{abstract}
We theoretically study the effect of magnetic moir\'e superlattice on the topological surface states by introducing a continuum model of Dirac electrons with a single Dirac cone moving in the time-reversal symmetry breaking periodic pontential. The Zeeman-type moir\'e potentials generically gap out the moir\'e surface Dirac cones and give rise to isolated flat Chern minibands with Chern number $\pm1$. This result provides a promising platform for realizing the time-reversal breaking correlated topological phases. In a $C_6$ periodic potential, when the scalar $U_0$ and Zeeman $\Delta_1$ moir\'e potential strengths are equal to each other, we find that energetically the first three bands of $\Gamma$-valley moir\'e surface electrons are non-degenerate and realize i) an $s$-orbital model on a honeycomb lattice, ii) a degenerate $p_x,p_y$-orbitals model on a honeycomb lattice, and iii) a hybridized $sd^2$-orbital model on a kagome lattice, where moir\'e surface Dirac cones in these bands emerge. When $U_0\neq\Delta_1$, the difference between the two moir\'e potential serves as an effective spin-orbit coupling and opens a topological gap in the emergent moir\'e surface Dirac cones.
\end{abstract}

\date{\today}


\maketitle

\emph{Introduction.-}Recently, mori\'e superlattices in twisted two-dimensional (2D) materials provide a novel platform to study a variety of strong correlation effects in flat minibands. Two prime examples are twisted graphene and transition metal dichalcogenide (TMD) multilayers~\cite{andrei2020,balents2020,carr2020,bistritzer2011,cao2018}. Motivated by the success of twisted van der Waals heterostructures, it is natural to study the effect of moir\'e superlattice of the Dirac cone on the surface of a 3D topological insulator (TI). Moir\'e superlattices in TI materials are ubiquitous, either in TI film grown on lattice mismatched substrate, or misalignment of topmost quintuple layer in bulk Bi$_2$Te$_3$~\cite{song2010,wang2012,jeon2011,liu2014,xus2015,xuj2015,schouteden2016,kar2019}. Previous studies have revealed the folded gapless Dirac cone within the bulk gap due to its topological nature from the time-reversal (TR) invariant mori\'e superlattice, where moir\'e surface states do not form isolated minibands~\cite{vargase2017,wangt2021,cano2021}. Thus to introduce TR breaking is a natural step to obtain isolated and even topological moir\'e surface minibands. The incorporation of the magnetic proximity effect into topological surface states have significantly enriched the variety of quantum matter~\cite{hasan2010,qi2011,tokura2019} exemplified in heterostructures with magnetic insulators~\cite{wei2013,katmis2016,tang2017} and intrinsic magnetic TI~\cite{zhang2019,otrokov2019,li2019,gong2019,wuj2019,yan2020,hu2020}, in particular van der Waals MnBi$_2$Te$_4$, which is compatible with the Bi$_2$Te$_3$ family materials~\cite{rienks2019}. Therefore it is straightforward to twist the van der Waals heterostructure of TI and magnetic insulator, while their effect on topological surface states have not been studied theoretically.

In this paper, we study the band structure of magnetic moir\'e surface states of TIs. Until now, all of the experimental moir\'e systems are TR invariant at the single particle level, thus the total Chern number is always equal to zero. Therefore, even with flat bands, it is quite difficult to achieve TR breaking interacting topological states such as the fractional Chern insulator in these systems. This motivates us to consider moir\'e superlattice of magnetic topological surface states. The topological nature of moir\'e surface Dirac cones is protected by TR symmetry, we find a Zeeman-type moir\'e potential generically opens the gap in the moir\'e surface Dirac cones and give rise to isolated flat Chern minibands with Chern number $\pm1$. In a $C_6$ periodic potential, the $\Gamma$-valley moir\'e surface electrons simulate 2D honeycomb lattice physics, leading to emergent moir\'e surface Dirac cones.

\begin{figure*}[t]
\begin{center}
\includegraphics[width=6.0in,clip=true]{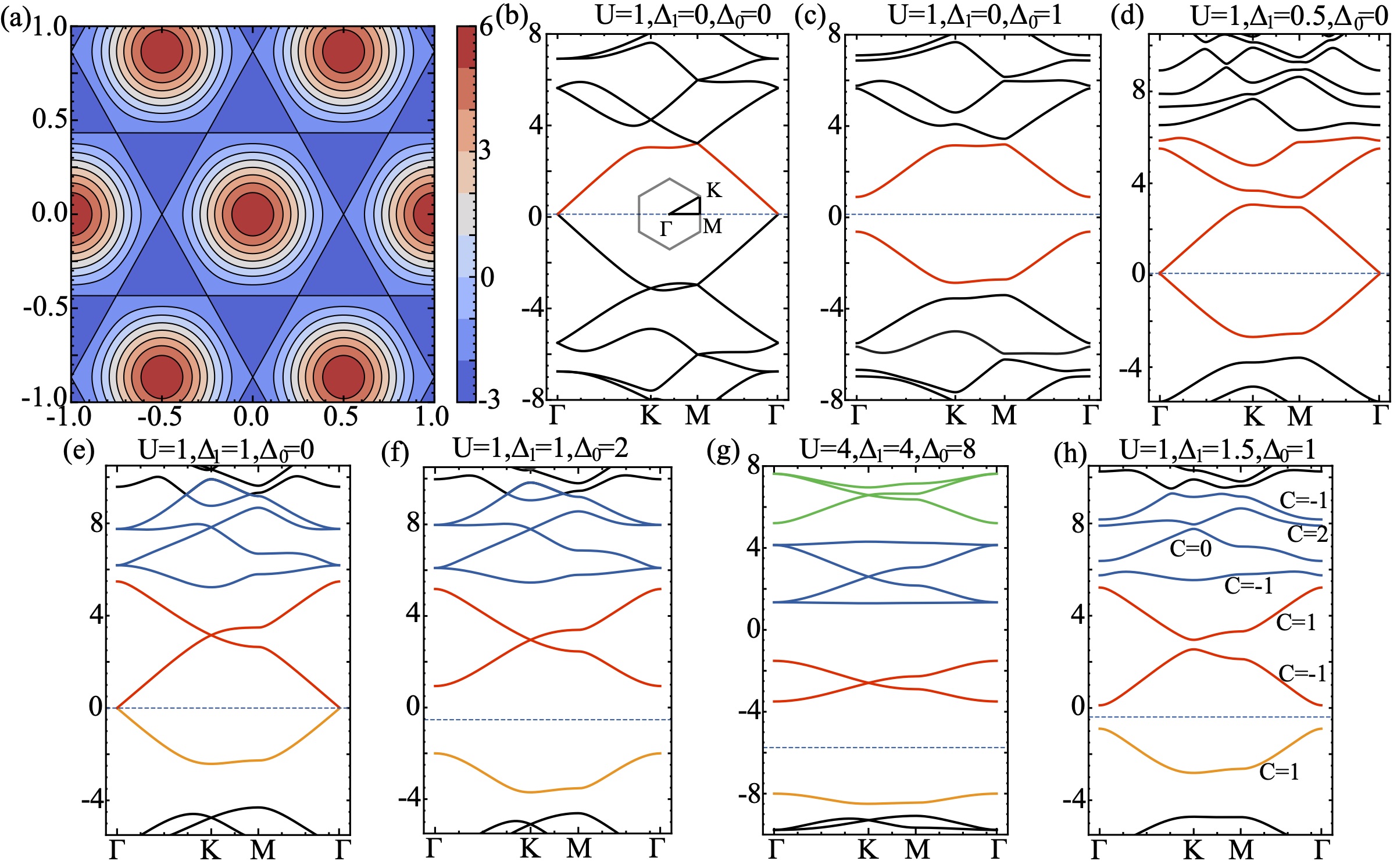}
\end{center}
\caption{(a) Schematic diagram of a $C_6$ moir\'e potential. (b) Energy spectrum at potential $(U_0,\Delta_1,\Delta_0)=(1,0,0)$. The corresponding MBZ is shown in the center. The entire spectrum remains gapless due to TR symmetry. The Dirac points are at $\Gamma$ and $M$.  The Zeeman moir\'e potential or uniform Zeeman term will generically gap out Dirac points as shown in (c,d) with $(U_0,\Delta_1,\Delta_0)=(1,0,1)$ and $(1,0.5,0)$, respectively. (e)-(g) When $U_0=\Delta_1$, the first three conduction bands originate from $s$ orbital, $p_x,p_y$ orbitals at honeycomb lattice and $sd^2$ orbtials at kagome site. All these band structure has characteristic Dirac points at high symmetric points despite of finite $\Delta_0,\Delta_1$.  (h) $U_0\neq\Delta_1$ effectively introduce spin-orbit coupling and introduces a topological gap, and each bands gain non-trivial Chern numbers. All parameters are in unit of $v_F/L$.}
\label{fig1}
\end{figure*}

\emph{Moir\'e Dirac electron.-}We introduce and study a model of TI surface Dirac fermions in periodic scalar potential with TR breaking and analyze its normal band structure and topology in mori\'e Brillouin zone (MBZ). Now we start with the massless Dirac fermion with a single Dirac cone at $\Gamma$ in 2D
\begin{equation}
\mathcal{H}_0(\mathbf{k})=v_F\left(k_x\sigma_y-k_y\sigma_x\right),
\end{equation} 
where $v_F$ is the Fermi velocity, $\boldsymbol{\sigma}=(\sigma_x,\sigma_y,\sigma_z)$ are Pauli matrices, $\mathbf{k}=(k_x,k_y)$ are the 2D momentum. A uniform exchange coupling between an out-of-plane magnetization and Dirac fermion opens a gap in the surface spectrum, and leads to the surface quantum Hall effect with a half-quantized Hall conductance. Now we put this Dirac fermion in a periodic TR breaking mori\'e potential $U(\mathbf{r})$ with a discrete translational symmetry, then to the lowest-order perturbation,
\begin{eqnarray}
\mathcal{H} &=& \mathcal{H}_0(-i\partial_\mathbf{r})+U(\mathbf{r}),
\\
U(\mathbf{r}) &=& U_0(\mathbf{r})\sigma_0+\Delta(\mathbf{r})\sigma_z,
\end{eqnarray}
where $U(\mathbf{r})=U(\mathbf{r}+\mathbf{L}_{1,2})$, and $\mathbf{L}_{1,2}$ are two primitive vectors of the mori\'e superlattice, $\sigma_0$ is the identity matrix. $U_0(\mathbf{r})$ is the scalar potential, and $\Delta(\mathbf{r})$ is the Zeeman-type potential from magnetic exchange interaction which contains both the moir\'e and uniform parts. This model can apply to bulk TI crystal with twisted surface states in the interface between a TI and a ferromagnetic insulator.

Although all of the physical effects discussed in this paper are generic for any magnetic moir\'e superlattice on topological surface states. To be concrete, we would like to start from a simple model describing the TIs of Bi$_2$Te$_3$ family~\cite{qi2011}, where threefold rotations with respect to the $z$-axis ($C_{3z}$) require that $U(\mathbf{r})=U(C_{3z}\mathbf{r})$. We consider the periodic potential with a form
\begin{equation}
U(\mathbf{r})=2\left(U_0+\Delta_1\sigma_z\right)\sum\limits_{j=1}^3\cos\left(\mathbf{G}_j\cdot\mathbf{r}\right)+\Delta_0\sigma_z,
\end{equation}
where $\mathbf{G}_j=(4\pi/\sqrt{3}L)\left(-\sin(2\pi j/3),\cos(2\pi j/3)\right)$ are three reciprocal vectors. $(U_0, \Delta_1,\Delta_0)$ are the mori\'e potential strength, which are scalar potential, Zeeman mori\'e potential, and uniform Zeeman term. There are two sets of energy scales $v_F/L$ and $(U_0, \Delta_1,\Delta_0)$ in this system, and the low-energy physics is determined by the ratio between $v_F/L$ and $(U_0, \Delta_1,\Delta_0)$. In particular, as we will see in the following, the interplay of moire potential $(U_0,\Delta_1,\Delta_0)$ would give rise to different band topology.

\emph{Band structure.-}The band structures are calculated by using the plane-wave expansion with a cutoff of 80 basis sets. First we estimate the energy scale in the system. For a typical TI such as Bi$_2$Te$_3$, the Dirac velocity is $v_F\approx250$~meV$\cdot$nm~\cite{chen2009,xia2009}. If we set the moir\'e lattice length as $L=10$~nm with the twisted angle $\theta=a/L$, then the energy scale of $v_F/L=25$~meV. Thus, in the unit of $v_F/L$, we expect energy scale of the effective potential at moir\'e scale is at the order of tens of meV~\cite{kerelsky2019,wu2019,zhang2020,angeli2021}, namely $\Delta_0$ is around $[0,2]$, $\Delta_1$ is $[0,2]$, and $U_0$ from $[0,2]$. 

Fig.~\ref{fig1}(a) is a schematic diagram of the $C_6$ moir\'e potential, and the minimum of potential constitutes a honeycomb lattice. We therefore expect the physics of the moir\'e band are generated by orbitals sitting on the honeycomb sites. In Fig.~\ref{fig1}(b) with $(U_0,\Delta_1,\Delta_0)=(1,0,0)$, the moir\'e potential $U_0$ folds the surface Dirac bands, and due to TR symmetry, the moir\'e surface Dirac cones at $\Gamma$ are preserved and satellite Dirac points at found at $M$ points, which is consistent with previous study~\cite{wangt2021,cano2021}. Moreoever, $U_0$ introduces the particle-hole asymmetry. Generically these minibands are non-degenerate, since they are from the spin-orbit coupled topological surface states. Now we study the effect of TR breaking potentials. By turning on a finite either Zeeman term $\Delta_0$ or moir\'e potential $\Delta_1$, all the moir\'e Dirac cones are generically gapped as shown in Fig.~\ref{fig1}(c) and Fig.~\ref{fig1}(d), respectively. 

\begin{figure}[b] 
\begin{center}
\includegraphics[width=3.4in,clip=true]{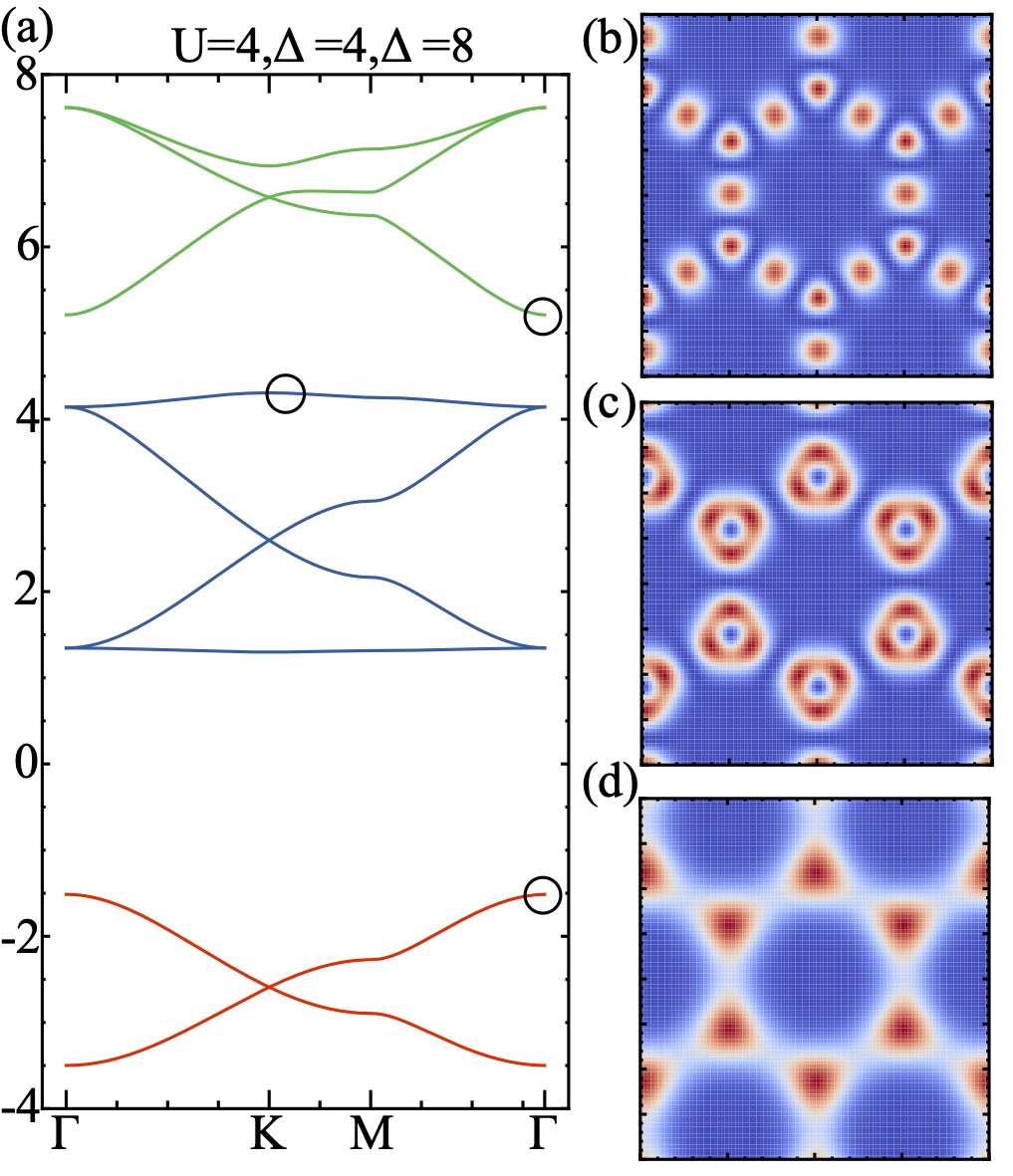}
\end{center} 
\caption{(a) Energy spectrum with $(U_0,\Delta_1,\Delta_0)=(4,4,8)$. The three sets of bands originate from $s$ Wannier orbitals on the honeycomb lattice, $p_x\pm ip_y$ Wannier orbitals on the honeycomb lattice, and hybridized $sd^2$ orbtials on the kgome lattice. (b)-(d) Wave function density distribution of three selected Bloch states encircled in (a), which clearly show the different orbital characters. Here we only present spin up part for the moir\'e potential only acts on spin up when $U=\Delta_1$.}
\label{fig2}
\end{figure} 

In particular, when $U_0=\Delta_1$, we find the re-emergence of the gapless moir\'e Dirac cones at $K$ in conduction bands of different energies as shown in Fig.~\ref{fig1}(e). As shown more clearer in Fig.~\ref{fig1}(h), the first three conduction bands can be classified into three different groups. To reveal the nature of moir\'e band physics, we identify the symmetries and centers of the Wannier orbitals underlying the moir\'e bands by employ topological quantum chemistry~\cite{bradlyn2017}. We first compute the symmetry of the Bloch states and classifying them in terms of the irreducible representations of the little groups at the corresponding high symmetry points, and then compare the list of irreducible representations with the Elementary Band Representations (EBR) of the space group $P6mm$ listed on the Bilbao Crystallographic server~\cite{bilbao1,bilbao2,bilbao3,elcoro2017}. The results are listed in Fig.~\ref{fig2}. Consistent with the emergent honeycomb structure of the moir\'e potential, the first set of bands is formed by $s$ orbital on the honeycomb lattice. These bands form a Dirac point at $K$ and are topologically equivalent to the $\pi$ bands of graphene except there is no spin degeneracy. The second set of bands, is formed by $p_x\pm ip_y$ orbitals on an honeycomb~\cite{wu2007,wang2021} that form a pair of almost dispersionless bands and also have a Dirac node at $K$. The third set of bands is formed by three bands which has one flat band and Dirac point at $K$. The symmetry analysis reveals that they are generated by orbitals centered on a kagome lattice, namely on the 3c Wyckoff positions lie at the mid bonds between two honeycomb sites. Such orbitals comes from hybridization of two different orbitals at honeycomb sites, which  effectively transforms the hexagonal symmetry of the moir\'e lattice into the physics of the kagome lattice, namely the $sd^2$ graphene~\cite{zhou2014}. It is worth mentioning that similar results found in $\Gamma$-valley TMD moir\'e bands~\cite{angeli2021}. However, we point out there is an essential \emph{difference} is that here the moir\'e bands are non-degenerate, and the Dirac points are emergent even in the presence of TR breaking at $U_0=\Delta_1$ and are protected by $C_{3v}$. As we see in the following, when $\Delta_1$ deviates from $U_0$, the deviation effectively act as the spin-orbit coupling and opens a topological gap at Dirac points. While in $\Gamma$-valley TMD moir\'e bands, the Zeeman potential will not lead to gap opening but just split the spin up and down bands in energy. It is noted that for an opposite Zeeman potential, all similar physics will occur on highest in energy moir\'e valence bands when $U_0=-\Delta_1$.

\begin{figure}[t]  
\begin{center}
\includegraphics[width=3.4in,clip=true]{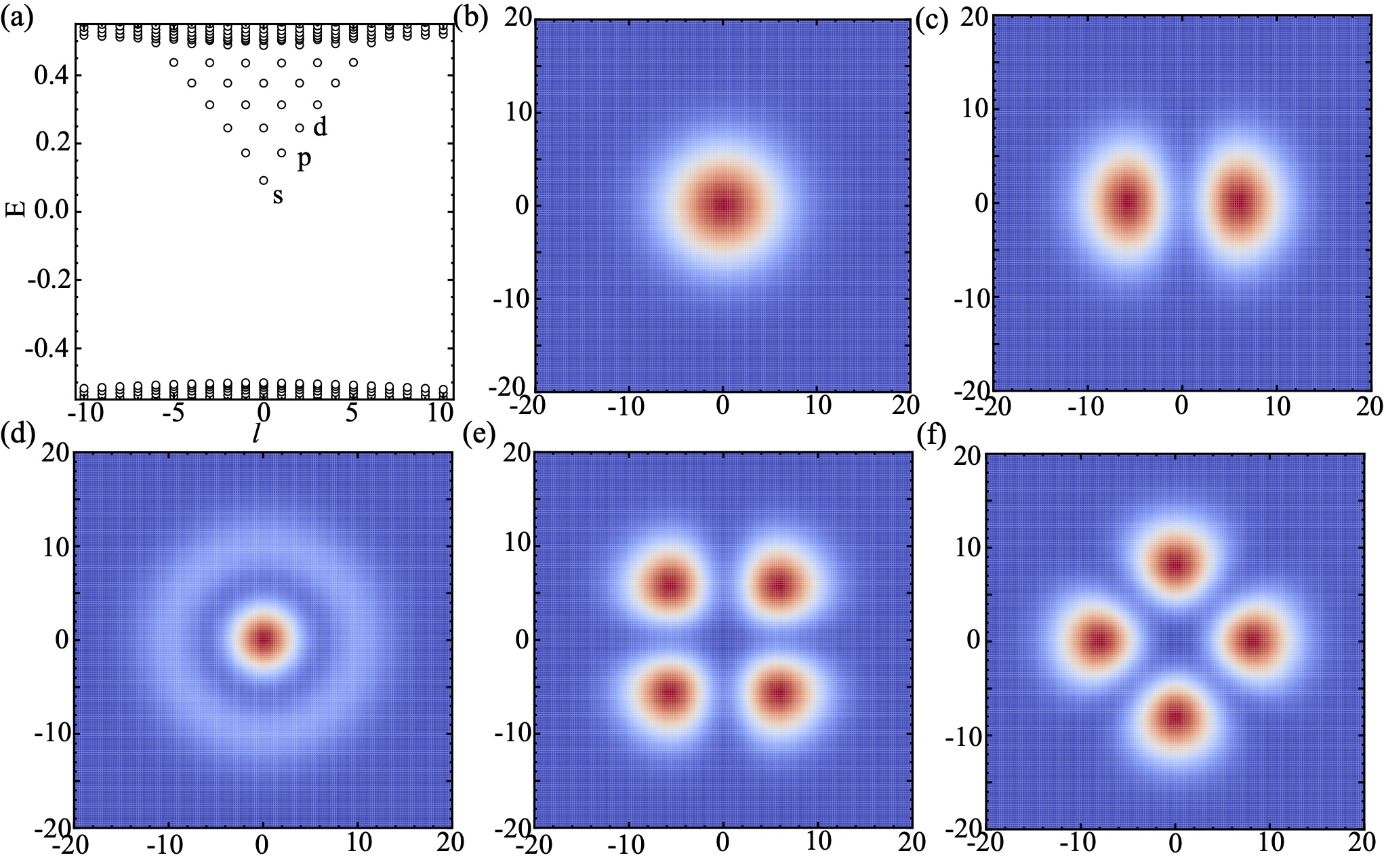}
\end{center} 
\caption{(a) The bound states spectrum of Dirac fermion in the moir\'e potential valley. (b)-(f) The wave function density distributions of the bound states in energetic order show the $s$, $p_x$,$d_z^2$, $d_{xy}$, $d_{x^2-y^2}$ orbitals characteristics. $(p_x\pm ip_y)$, $(d_{xy}\pm id_{x^2-y^2})$ are degenerate.}
\label{fig3}
\end{figure}

To illustrate how these orbitals emerges in the above moir\'e bands when $U_0=\Delta_1$, we analyze the spectrum of Dirac fermion in a rotational invariant potential, namely $\mathcal{H}_0(-i\partial_{\mathbf{r}})+\Delta_0\sigma_z+\Delta_1V(\mathbf{r})(\sigma_0+\sigma_z)$. Thus the moir\'e potential only acts on spin up. We approximate the moir\'e potential near its minimum as a harmonic trap, $V(\mathbf{r})\propto (\mathbf{r}/L)^2$ with a cutoff length of $R_0\approx 0.2L$. Fig.~\ref{fig3}(a) shows the energy spectrum from the numerical calculation performed on a disc geometry with radius $R=4R_0$. There are bound states in the gap
which are classified into $s,p$, and $d$ orbitals in energetic order, where their wave function density distributions for corresponding bound states are clearly demonstrated in Fig.~\ref{fig2}(b)-(f). Thus the tight-binding model from hopping of the bound states at the honeycomb lattice naturally give rise to the above moir\'e bands. 

Then we study the physics away from $U_0=\Delta_1$. A typical spectrum is shown in Fig.~\ref{fig1}(h) where the emergent Dirac points (as in Fig.~\ref{fig1}(e-g)) are gapped, and one gets isolated flat bands. Such a gap opening must be topological as we can expect from EBR. We further calculate Chern number of the minibands which is well defined here, for the moir\'e potential regularizes the Dirac fermion into MBZ. As expected, these bands feature nontrivial Chern numbers. Furthermore, the $p_x\pm ip_y$ and $d_{xy}\pm id_{x^2-y^2}$ bound states are no longer degenerate when $U_0\neq\Delta_1$.  Therefore, the magnetic moir\'e potential $\delta=U_0-\Delta_1$ (deviated from $U_0=\Delta_1$) is an effective spin-orbit coupling and act as a topological mass term in the emergent Dirac bands~\cite{wang2021}, and the sign of Chern numbers is determined by the sign of $\delta$. The bandwidth of the flat Chern band from $p_x,p_y$ orbitals is about $\mathcal{W}\approx0.2v_F/L$. Interestingly, the first pair of conduction bands essentially realize the Haldane mode on the honeycomb lattice~\cite{haldane1988}, where the bandwidth is on the order of $\mathcal{W}\approx v_F/L$ and tuned by moir\'e lattice constant. 

\begin{figure}[b]  
\begin{center}
\includegraphics[width=3.4in,clip=true]{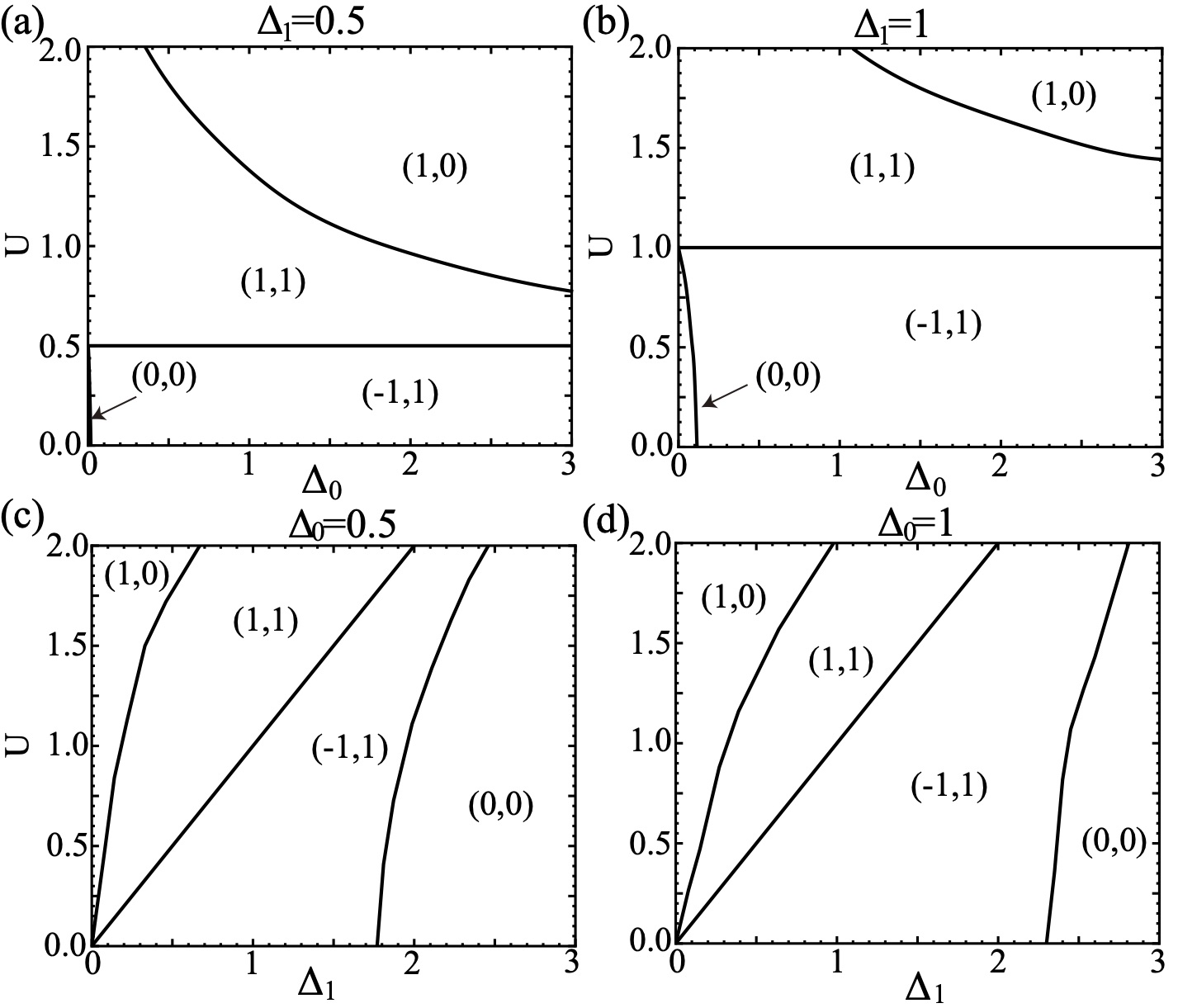}
\end{center}     
\caption{Chern number phase diagram of the first conduction and valance bands $(C_c,C_v)$ as function of $(U,\Delta_0)$ with $\Delta_1=0.5$ in (a) and $\Delta_1=1$ in (b). (c,d) $(C_c,C_v)$ as function of $(U,\Delta_1)$ for $\Delta_0=0.5$ and $\Delta_0=1$, respectively.}
\label{fig4}
\end{figure} 

Now we understand the physics in magnetic moir\'e surface spectrum and band topology. We further calculate the Chern number phase diagram of the first conduction and valence bands as shown in Fig.~\ref{fig4}. The phase boundary is obtained by finding the gap closing points. With a fixed $\Delta_0$, $(C_c,C_v)=(1,0)$ when $\Delta_1$ is small. The valence band will acquire a finite Chern number $C_v=1$ from remote bands by increasing $\Delta_1$. When $\Delta_1$ further increases, the conduction band will exchange Chern number with higher bands at $K$ on $U=\Delta_1$ line and get $C_c=-1$. With further increasing $\Delta_1$, the conduction and valence bands become topological trivial by exchange Chern number with each other at $\Gamma$. As shown in Fig.~\ref{fig4}(a,b) with a fixed $\Delta_1$, when $\Delta_0$ is relatively small, the valance and conduction bands are always trivial when $U_0<\Delta_1$. Similarly, as $\Delta_0$ increases, the conduction and valence bands exchange Chern number with each other at $\Gamma$ and becomes $(C_c,C_v)=(-1,1)$. Now by crossing the $U_0=\Delta_1$ line, the conduction band has  $C_c=1$ which exchanges Chern number at $K$ with higher band. More results on the evolution of phase diagram are illustrated in Supplementary Materials~\cite{supp}.

\emph{Discussion.-} The bandwidth of the flat Chern bands of moir\'e surface states is tuned by the twisting angle and is significantly smaller than the Coulomb repulsion energy, which make it an ideal platform for realizing interacting topological states~\cite{levin2009,maciejko2010,qi2011f,tang2011,sun2011,neupert2011,spanton2018}. For an estimation, taking the dielectric constant of TI surface states $\epsilon_r\approx5$, one obtains a Coulomb interaction energy $\mathcal{U}=e^2/\epsilon_rL\approx30$~meV. Thus $\mathcal{U}/\mathcal{W}\gtrsim1$ for filling in the first sets of conduction/valence bands, while $\mathcal{U}/\mathcal{W}\gtrsim6$ for filling in the second sets of conduction band. Furthermore, even with either Chern number $0$ or $\pm1$, the nondegenerate flat band allows a single Fermi surface with large density of states when partially filled, leading to a chance of realizing an intrinsic TR breaking superconductivity.

The band topology of lowest in energy magnetic moir\'e conduction bands is essentially rooted in the $C_6$ periodic potential for a simple realization of honeycomb lattice, which is the case for Bi$_2$Te$_3$. In a $D_4$ periodic potential, one could also get flat Chern band but these interesting emergent Dirac cone from orbital-active models in biparticle lattice will not occur. Here we emphasize the results are different from previous study on twisted magnetic TI bilayer, there strong hybridization between top and bottom gapped surface states occurs~\cite{lian2020}.

Next we briefly discuss the superconducting proximity effect. Without the moir\'e potentials, the proximitized surface state is always a topological superconductor when $\Delta_0<\Delta_s$ if $\mu=0$ or $\mu>\Delta_0$, with $\Delta_s$ is the $s$-wave pairing amplitude. We find by adiabatically turning on the moir\'e potential, the band structure changes but without gap closing. Namely, the system is always in the topological superconducting state with above condition~\cite{supp}. Therefore, Majorana zero mode in the vortex core and chiral Majorana edge modes are expected~\cite{fu2008,wang2015c}.

Finally, we discuss the feasibility to realize our model of magnetic Dirac fermion in periodic potential. Mechanically robust single septuple layer of MnBi$_2$Te$_4$ has been obtained experimentally~\cite{deng2020}, making it possible to implement twisted superlattice on Bi$_2$Te$_3$ surfaces. The wave function density of Dirac surface state resides in both MnBi$_2$Te$_4$ layer and topmost layer of Bi$_2$Te$_3$~\cite{otrokov2017}, therefore both $\Delta_0$ and $\Delta_1$ are present. With the exchange coupling at the order of hundreds of meV, we expect the magnetic potential at moir\'e scale is at the order of tens of meV, namely $(U_0,\Delta_1,\Delta_0)$ is at the same order.

\emph{Summary.-}We find the magnetic potentials generically gap out the moir\'e surface Dirac cones and lead to isolated flat Chern minibands with Chern number $\pm1$, thus the fractional filling in it makes the strongly correlated topological states possible. The moir\'e surface electrons in a $C_6$ periodic potential simulate the physics of orbital-active honeycomb lattice, and the magnetic moir\'e potential acts as an effective spin-orbit coupling. Our model provides a convenient condensed matter platform to engineer the Haldane model with narrow bandwidth and strong interaction.

\begin{acknowledgments}
We acknowledge Y. Zhang for stimulating discussions. This work is supported by the National Key Research Program of China under Grant Nos.~2019YFA0308404 and 2016YFA0300703, the Natural Science Foundation of China through Grant No.~11774065, Shanghai Municipal Science and Technology Major Project under Grant No.~2019SHZDZX01, Science and Technology Commission of Shanghai Municipality under Grant No.~20JC1415900, and the Natural Science Foundation of Shanghai under Grant No.~19ZR1471400.
\end{acknowledgments}

\end{document}